\documentclass[pdftex]{aa}
\usepackage{graphicx}
\usepackage{txfonts}
\begin{document}

\title{Doppler Probe of Accretion onto a T Tauri star
          \thanks{Based on observations collected at the European Southern Observatory 
					(75.C-0292).}}


   \author{ P. P. Petrov\inst{1}
                   \and G. F. Gahm\inst{2}
                       \and G. J. Herczeg\inst{3}
                      \and H. C. Stempels\inst{4}
                                      \and F. M. Walter\inst{5}
          }
   \offprints{P. P. Petrov}

      \institute{Crimean Astrophysical Observatory, Taras Shevchenko National University
              of Kiev, 98409 Nauchny, Crimea,  Ukraine\\
              email: \mbox{petrov@crao.crimea.ua}
       \and Stockholm Observatory, AlbaNova University Centre, Stockholm University, SE-106 91 Stockholm, Sweden
	      \and The Kavli Institute for Astronomy and Astrophysics, Peking University, Yi He Yuan Lu 5, HaiDian Qu, Beijing 100871, P.R. China 
	      	     \and Department of Physics and Astronomy, Uppsala University, Box 516, SE-75120 Uppsala, Sweden 
	           \and Department of Physics and Astronomy, Stony Brook University, Stony Brook, NY 11794-3800, USA 
      }
   \date{}

   
\abstract
	 {The YY Ori stars are T Tauri stars with prominent time-variable redshifted absorption components that flank certain emission lines. S CrA, one of the brightest of these stars, affords the rare opportunity of directly probing the accretion processes on the line of sight to one of the components of this wide visual pair.}
   {We followed the spectral changes occurring in S CrA to derive the physical structure of the accreting gas.}
   {A series of high-resolution spectra of the two components of S CrA was obtained during four nights with the UVES spectrograph at the Very Large Telescope.}
   {We found that both stars are very similar with regard to surface temperature, radius, and mass. Variable redshifted absorption components are particularly prominent in the SE component.  During one night, this star developed a spectrum unique among the T Tauri stars: extremely strong and broad redshifted absorption components appeared in many lines of neutral and ionized metals, in addition to those of hydrogen and helium. The absorption depths of cooler, low ionization lines peak at low velocities - while more highly ionized lines have peak absorption depths at high velocities.  The different line profiles indicate that the temperature and density of the accretion stream increase as material approaches the star.  We derive the physical conditions of the flow at several points along the accretion funnel directly from the spectrum of the infalling gas. We estimated mass accretion rates of about 10$^{-7}$~M$_\odot$/yr, which is similar to that derived from the relation based on the strength of H$\alpha$ emission line. }
  {This is the first time the density and temperature distributions in accretion flows around a T Tauri star have been inferred from observations. Compared with predictions from standard models of accretion in T Tauri stars, which assume a dipole stellar magnetic field, we obtained higher densities and a steeper temperature rise toward the star.}


 \maketitle


\section{Introduction}

Some classical T Tauri stars (CTTs) occasionally show broad redshifted absorption components that flank certain emission lines that indicates infalling gas at velocities of $>$~100~km\,s$^{-1}$. These redshifted absorptions (inverse P Cygni profiles) were first identified in CTTs by Walker (\cite{Walker72}), who introduced the class of YY Orionis stars, which exhibit these features.  Stars that have the inverse P Cygni  profiles also in lines of neutral and singly ionized elements in addition to the hydrogen lines  are very rare.
 
One of the most prominent YY Ori stars with redshifted metallic absorptions is the $1\farcs3$ binary S CrA at a distance of about 130 pc (see Appenzeller et al. \cite{Appenzeller86}, and references therein).  Carmona et al. (\cite{Carmona07}) spatially resolved the binary and identified the SE component as exhibiting the strongest  inverse P Cygni features.  Both components vary by about two magnitudes in $V$ (Walter \& Miner \cite{Walter05}). On average, the NW component is brighter by about 0.4$^m$, but either component can become the brightest in the system. The two S CrA stars have optical spectra covered in bright metal lines (Gahm et al.~2008), which is a common trait of CTTs with high accretion rates (Hamann \& Persson \cite{Hamann92}; Beristain et al. \cite{Beristain98}).  CO fundamental lines from S CrA SE have broad profiles, which suggests an inner disk radius of 0.1 $\sin^2$ i AU (Bast et al. \cite{Bast11}, Brown et al. \cite{Brown13}), or $\sim 4$ R$_\ast$ with an inclination of $\sim 30^\circ$ (Schegerer et al. \cite{Schegerer09}).

We obtained high-resolution spectra of each of the components in S~CrA during four consequetive nights. In a previous study (Gahm et al. \cite{Gahm08}) we analyzed the strength and variability of veiling in both stars. In this Letter we focus on the inverse P Cygni spectrum in S~CrA~SE, which can become exceptionally rich on occasion, and report on the results obtained from one such event in this, so far, unique T Tauri star.    

\section{Observations}

Eleven high-resolution spectra of S CrA were obtained between August 14\,--\,17, 2005 with the UVES spectrograph in the dichroic-mode on the 8 m VLT/UT2 of the European Southern Observatory, Chile. The stars were aligned in the slit and were perfectly spatially resolved, therefore the two components were extracted individually.  The spectra have a high signal-to-noise ratio ($\ga 100$), a spectral resolution of $R \approx 60\,000$, and a wavelength coverage from 3500 to 6700 {\AA}. For more information on data reductions see Stempels et al. (\cite{Stempels07}). 

\section{Shapes and strengths of inverse P Cygni profiles}
\label{sec:results}

Both S~CrA NW and SE have very rich emission line spectra, which are similar on the whole. The main difference is that SE sometimes shows extremely strong and broad redshifted absorption components in many lines. Because of veiling, the photospheric absorption lines of both stars are weaker than those of normal stars of the same spectral type, but consistent with spectral type K7 for each component, contrary to the earlier estimations of G6 for primary and K5 for secondary, which were based on low-resolution spectra (Carmona et al.~\cite{Carmona07}). By matching our spectra with synthetic templates and comparing them with evolutionary tracks by Siess et al.~(\cite{Siess00}), we assigned for each component $T_{eff}$ = 4250~K, $L_*$ = 0.7~L$_{\odot}$, $M_*$ = 1~M$_{\odot}$, $R_*$ = 1.4~R$_{\odot}$.  These values are consistent with the observed value of log g = 4.0.  

\begin{figure}
\centerline{\resizebox{8cm}{!}{\includegraphics{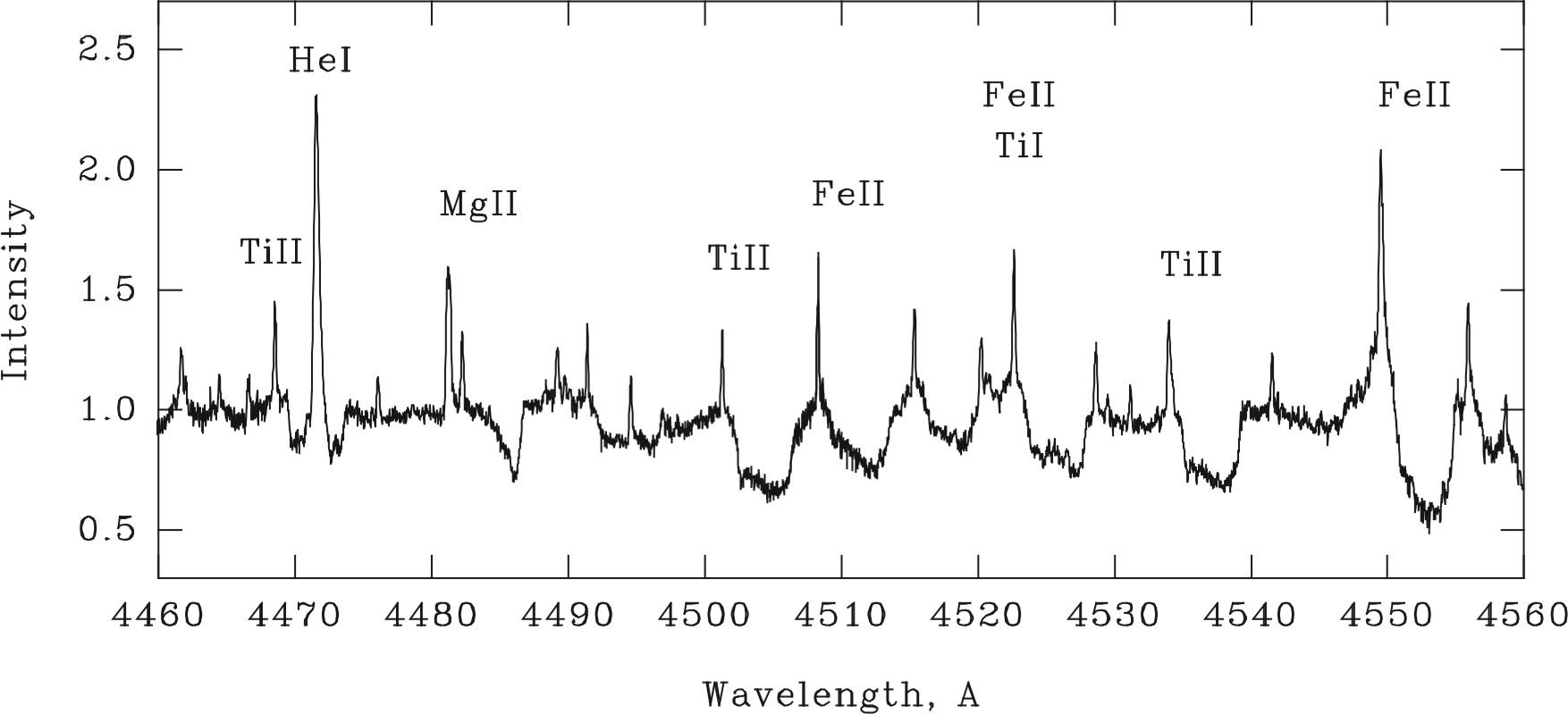}}}
\caption{Spectrum of S~CrA~SE  from August 14, 2005. 
Note the numerous narrow emission lines and associated redshifted absorption components.}
\label{4460_em}
\end{figure}

The veiling is quantified by the veiling factor (VF), which is defined as the ratio of an assumed continuum in excess to the stellar continuum. In the four nights of 2005 the lowest veiling (around $\lambda$\,5500\,\AA) in S CrA SE occurred on August 16,  when $VF\,\approx\,2$, and the highest veiling was observed on August 14  with $VF\,\approx\,4.5$. On this latter date (MJD 24553597.035), prominent inverse P Cygni profiles are present in many lines of the S~CrA SE spectrum obtained on August 14. On the next day, these features were still present, but less intense. Figures~\ref{4460_em} and 2 show the remarkable differences between lines of different elements, energies, and ionizations stages.  

Fig.~\ref{acr3} shows the inverse P Cygni profiles of three lines of different ionization and excitation states. It shows that lines formed at low temperatures are strongest at low infall velocity, while the reverse is true for the high-temperature lines, which absorb preferentially at higher redshifts.  The infall absorption in the low-ionization \ion{Ca}{i} line is detected at $\approx70$ km\,s$^{-1}$, peaks at 100 km\,s$^{-1}$, and falls off to zero at $\approx$350 km\,s$^{-1}$.  In contrast, the \ion{Fe}{ii} line peaks at 250 km\,s$^{-1}$. The \ion{Si}{ii} line, with a higher ionization potential, gradually increases until $\sim 330$ km\,s$^{-1}$. The red edge of the $\ion{Si}{ii}$ absorption is at about 380 km\,s$^{-1}$, which can be compared with the expected free-fall velocity of 470 km\,s$^{-1}$ at the stellar surface. We note that Bertout et al. (\cite{Bertout82}) found very similar correlations for S CrA in that the velocity shifts of absorption components in certain metal lines increase with the sum of the ionization and excitation potentials in the velocity range from 190 to 250 km\,s$^{-1}$. 

\begin{figure}
\centerline{\resizebox{9cm}{!}{\includegraphics{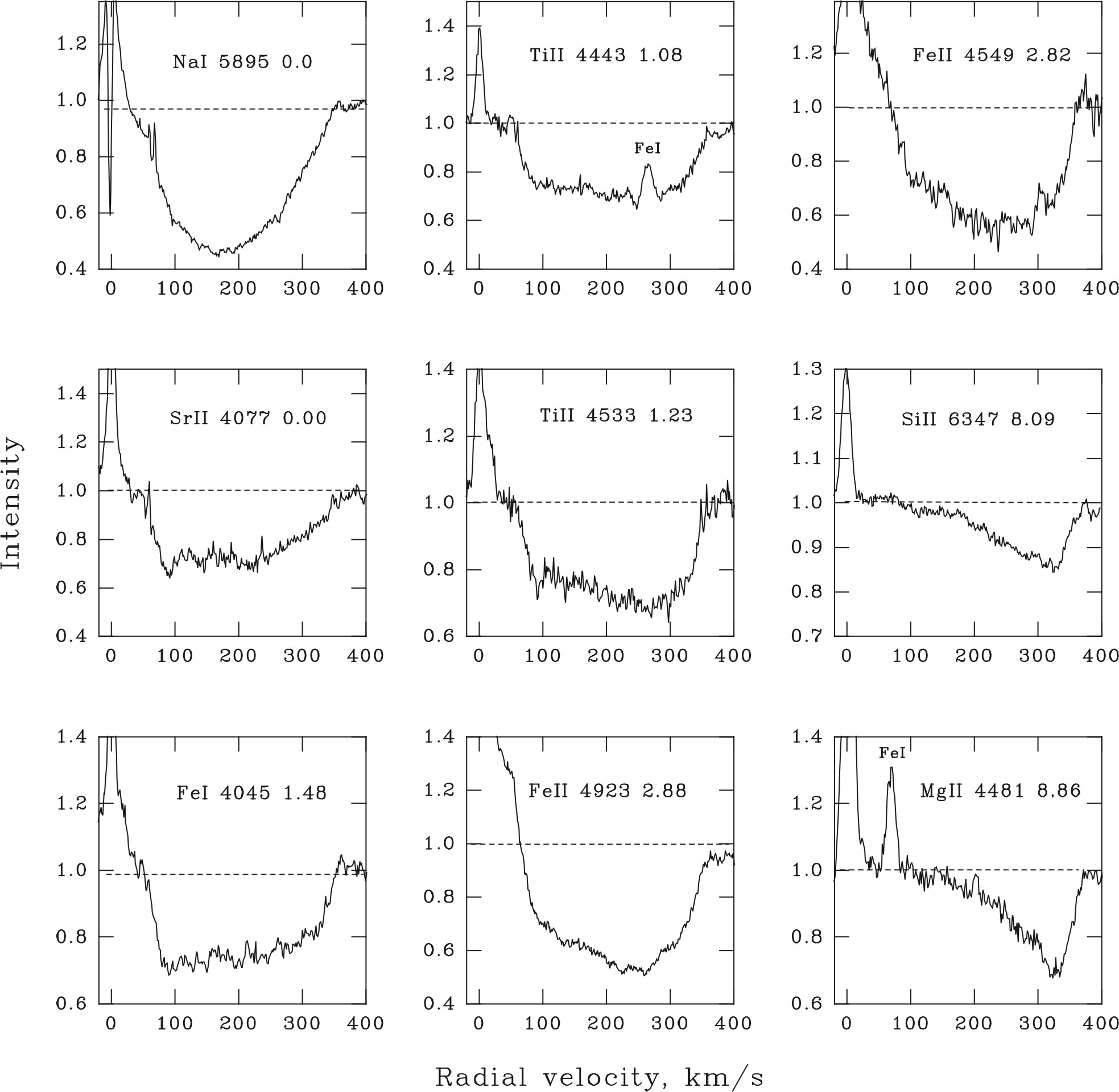}}}
\caption{Examples of inverse P Cygni profiles for different elements in S CrA SE. The wavelength (\AA) and excitation potential (eV) are indicated.}
\label{acr9}
\end{figure}  
 
\begin{figure}
\centerline{\resizebox{5cm}{!}{\includegraphics{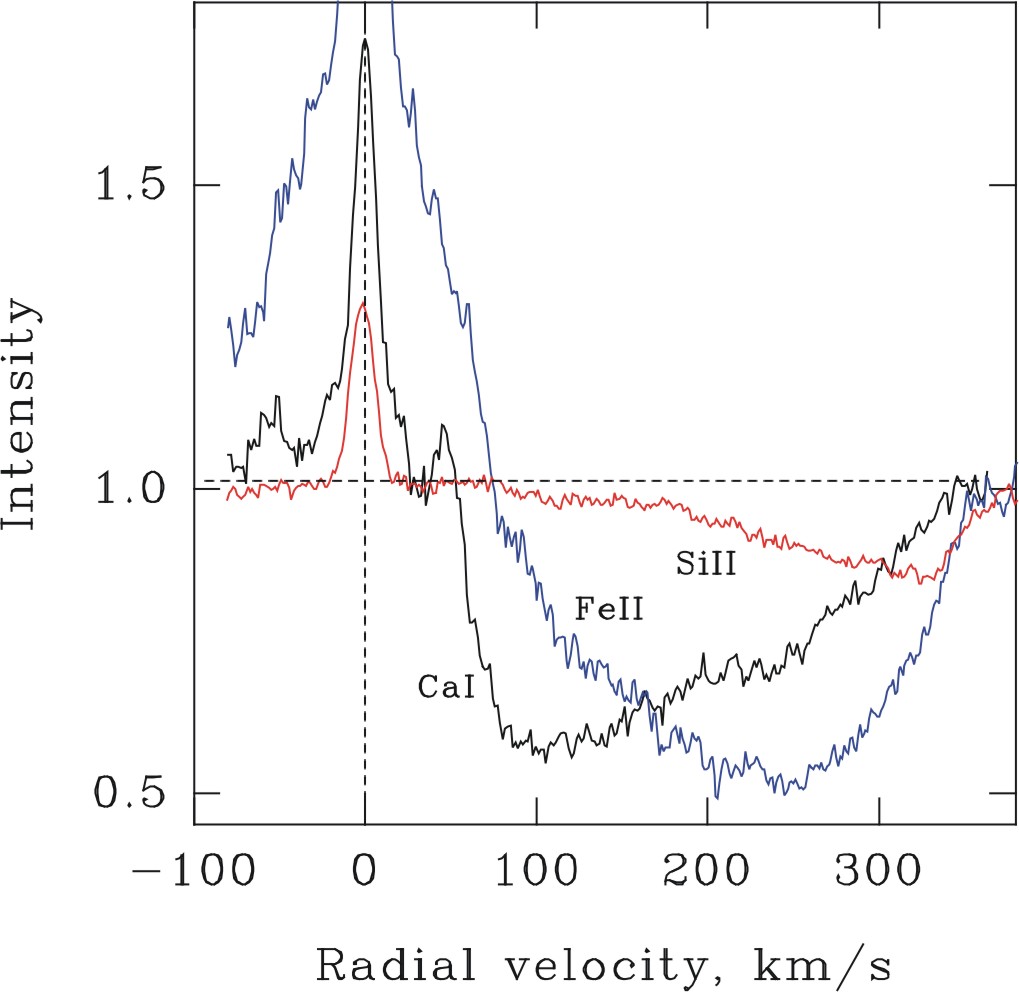}}}
\caption{Inverse P Cygni profiles in S CrA SE of lines of different excitation potentials: \ion{Ca}{i} 4226, 0.0 eV; \ion{Fe}{ii} 5018, 2.8 eV; \ion{Si}{ii} 6347, 8.1 eV 
(lower levels indicated). }
\label{acr3}
\end{figure} 
 
These differences establish that the physical conditions change along the accretion flow, which opens a rare opportunity to study  the structure and physical conditions in the accretion flow.The temperature and density profiles along the accretion flows have not been inferred from observations before, and are known  only as parameters derived from standard models of magnetospheric accretion (e.g., Hartmann et al.~1994; Martin~\cite{Martin96}; Muzerolle et al.~\cite{Muzerolle01}; Kurosawa et al. \cite{Kurosawa06}). Here, we estimate these parameters directly from the spectra of the infalling gas. 

Most of the inverse P Cygni absorptions of metals shown in Fig.2 are optically thin because of the steep velocity gradient in the accretion flow. The low optical depth is evident from comparing the line strengths within the same multiplet: the equivalent width of inverse P Cygni absorption is linearly proportional to the log $gf$ transition value. This means that the observed ratios of the line strenghs at a given velocity shift are defined only by the gas parameters and do not depend on the stellar surface area covered by the stream.

\section{Analysis: the structure of the accretion flow}

Under typical magnetospheric conditions the metals are expected to be collisionally excited (Beristain et al.~1998), i.e. the excitation distribution is set by the kinetic temperature, $T_{kin}$, and is described by the Boltzmann equation.   For the high densities $N_e$ $\geq$ 10$^{10}$ cm$^{-3}$ expected in an accretion flow, $T_{kin}$ $\approx$ $T_e$ (Mihalas~\cite{Mihalas78}). We assume LTE conditions, and did not consider possible non-LTE effects caused by UV radiation from an accretion shock.

For optically thin lines, the line strength should be proportional to the oscillator strength $f$ of the transition and the number of absorbing atoms  $N_s$ in the line of sight. In the following analysis we used the ratio of the line intensities, which is not affected by veiling for lines located in a restricted wavelength range. 

We consider a section of the accretion funnel restricted by a narrow range of infall velocities $dv$. The funnel flow can be represented by a slab of isothermal gas of solar chemical abundance, with an electron temperature $T_e$ and an electron density $N_e$. 

In an accretion flow with a high velocity gradient, the inverse P Cygni absorption at any velocity shift $v~-~v_0$~=~c~$\Delta\lambda$/$\lambda_0$ \noindent is formed in a thin layer of gas that moves with velocity $v$ toward the star. The geometrical thickness of this layer and the number of absorbing atoms depend on the velocity gradient. In the Sobolev approximation the optical depth in a line is

\begin{equation}
      $$\tau_v = \frac{\sqrt{\pi}e^2}{mc}\,f\,\frac{\lambda_0}{|dv/dr|}\,\,N_s$$,         
\end{equation}   

where $N_s$ is the number of absorbing atoms per cm$^3$ and $dv/dr$ (s$^{-1}$) is the velocity gradient  at distance $r$ from the stellar center, and the infall velocity is $v$ (cm\,s$^{-1}$). The gradient $dv/dr$ is defined by the free-fall equation:

\begin{equation}
       $$v(r) = \sqrt(2GM_*(\frac{1}{r} - \frac{1}{R_{ini}}))$$,  
\label{fall}            
\end{equation}

where $R_{ini}$ is the initial radius where the disk is truncated and gas is at zero velocity with respect to the star. This truncation radius is typically slightly smaller than the co-rotation radius, which depends on stellar mass $M_*$ and angular velocity $\Omega$:

\begin{equation}
      $ $r_{co}^3$ = G$M_*/\Omega^2$.  $
\label{corotation}            
\end{equation}

The rotational period of S CrA SE is not known.  Petrov et al. (\cite{Petrov11}) found a smooth change of about 5 km\,s$^{-1}$  in the radial velocity of the narrow emission component in the $\ion{He}{i}$  5875\,\AA\ line from August 14 to 17.  This implies a lower limit of the period of about three days. On the other hand, from the observed projected rotational velocity $v\,\sin i$ = 12 km\,s$^{-1}$ and the stellar radius of 1.4 $R_\odot$ an upper limit of 4.6 days can be deduced. With these limits, $r_{co}$ can be within 6--8 $R_*$. The observed infall velocity $\approx$~380 km\,s$^{-1}$ sets a lower limit of $R_{ini}$ $\approx$ 4 $R_*$. In the following we assume $R_{ini}$ $\approx$ 5 $R_*$, which is a typical value for TTS.

Our goal is to find $T_e$ and $N_e$ from ratios of different line depths. We used our code in which input parameters $T_e$ and $N_e$ are interactively adjusted so that the calculated line ratios reproduce  the observed ratios. For given values of $T_e$  and $N_e$, the electron pressure $P_e$ (dyn\,cm$^{-2}$), gas pressure $P_{gas}$ (dyn\,cm$^{-2}$), and mass density $\rho$ ($g\,cm^{-3}$) are defined by the gas laws. Then, the number density of absorbing atoms $N_s$ (cm$^{-3}$) is calculated using the Saha and Boltzmann equations. The particle density is $N_{\rm H} \approx \rho$/$m_{\rm H}$, where $m_{\rm H}$ is the hydrogen 
atom mass.

The temperature can be found from lines of ionized metals with different excitation potentials of the lower levels, such as \ion{Fe}{ii} (2-3 eV), \ion{Si}{ii} (8.1 eV), and \ion{Mg}{ii} (8.8 eV). Under typical magnetospheric conditions these elements are in their dominant stages of ionization, and therefore the line ratios (e.g., \ion{Fe}{ii} 4923/\ion{Mg}{ii} 4481) are weakly dependent on electron density. When $T_e$ is set, $N_e$ can be found from the ratios of \ion{Fe}{i}/\ion{Fe}{ii}. 

\begin{table}
\caption{Gas parameters measured at different infall velocities in S~CrA~SE}
\label{mdot_tab}
\begin{tabular}{cccccc} \hline

v(r) & T$_e$ & log\,N$_e$ &  log\,$\rho$ & log\,$\dot{M_{acc}}/w$ \\
(km\,s$^{-1}$) & (K) & (cm$^{-3}$) & (g\,cm$^{-3}$) & (M$_\odot$/yr) \\

\hline
100 & 5300 & 10.55 & -10.69  & -7.02  \\
200 & 6000 & 11.40 & -10.56  & -6.59  \\
250 & 6600 & 12.10 & -10.24  & -6.18  \\
300 & 7000 & 12.40 & -10.25  & -6.11  \\
330 & 7500 & 12.75 & -10.20  & -6.01  \\
\hline
\end{tabular}
\end{table}
 
\begin{figure}
\centerline{\resizebox{7cm}{!}{\includegraphics{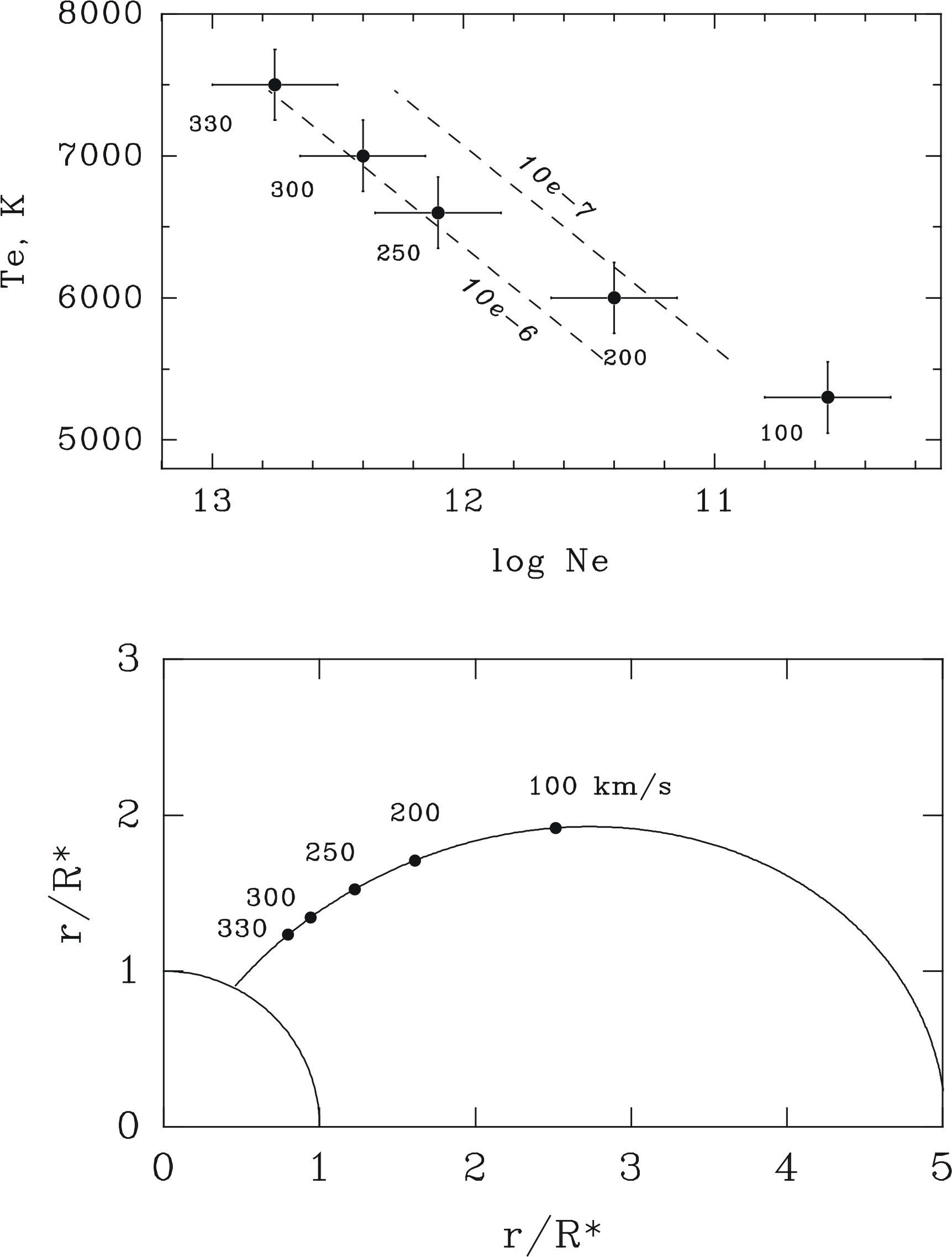}}}
\caption{Upper panel: parameters of the accreting gas (dots with error bars) at different infall velocities, as marked at the crosses. Dashed lines indicate loci of two constant accretion rates $\dot{M}_{acc}/w$\,\,in units of M$_\odot$/yr. Lower panel: a dipole magnetic loop connecting star and disk. The dots marked with infall velocities indicate positions at which the gas parameters were determined.
}
\label{ne_te}
\end{figure}  

After the appropriate pair of $T_e$, $N_e$ is found and the mass density $\rho$ is calculated, the mass accretion rate is defined as 

\begin{equation}
$  $\dot{M}_{acc} = \rho\,v\,A\,w$,  $
\label{accr_rho}
\end{equation}

where $A$(cm$^2$) is the stellar surface area ($4 \pi R_*^2$) and $w$ is the cross-section of the accretion funnel expressed as a fraction of the global stellar surface. Close to the stellar surface, $w$ should be equal to the size of the hot spot, typically about a few percent of the global stellar surface.  By the infall velocity we assume here the velocity vector radial to the stellar center, that is. only defined by gravity, while the velocity vector along the field lines may be different.

For the diagnostic of the infalling gas we measured lines of  the following multiplets: \ion{Fe}{i}\,(2, 4, 20, 42, 43), \ion{Fe}{ii}\,(27, 38, 42), \ion{Ti}{ii}\,(19, 50), and \ion{Mg}{ii}\,(4). The depths of the inverse P Cygni absorptions were measured at infall velocities of 100, 200, 250, 300, and 330 km\,s$^{-1}$. The stellar radial velocity of S~CrA SE is close to zero, and therefore the observed velocity shifts can be treated as infall velocities with respect to the star. The  oscillator strengths (log $gf$) were taken from the Vienna Atomic Lines Database (Kupka et al.~\cite{Kupka00}), and solar abundances from Lang\,(\cite{Lang74}). The results are given in Table \ref{mdot_tab}.

The upper panel of Fig.~\ref{ne_te} shows loci of parameters for the infalling gas on the $T_e-log\,N_e$ plane for different infall velocities. The error bars indicate the ranges of $T_e$ and $log\,N_e$ values that are obtained when considering the ratios of different pairs of spectral lines. All points fall between the two dashed lines of constant accretion rates, corresponding to $\dot{M}_{acc}/w$ = 10$^{-6}$ and 10$^{-7}$ M$_\odot$/yr. The lower values of $\dot{M}_{acc}/w$ at lower velocities are caused by the geometry of the magnetic funnel: the gas stream is wider farther out from the star. Note that the fraction $w$ enters into Eq.4.

Since the magnetic field lines are curved, the line of sight intersects different segments of the funnel flow as one approaches the star. This means that the temperatures and densities given in Table \ref{mdot_tab} do not describe a single field line, but probe the magnetosphere in  radial  direction to the star.  A schematic view of a dipole magnetic loop is shown in the lower panel of Fig.~\ref{ne_te}, where infall velocities and distances are calculated for $M_*$ = 1.0~M$_\odot$ and $R_{ini}$ $\approx$ 5~$R_*$.

The mass-accretion rate $\dot{M}_{acc}$ estimated from the model depends on the size of the hot spot.The filling factor (a fraction of the {\it global} stellar surface covered by the shock)   has been determined for many TTSs (e.g., Hartigan et al.~\cite{Hartigan91}; Valenti et al.~\cite{Valenti93}; Gullbring et al.~\cite{Gullbring00}). Typical values are about 1\%, but the range is wide, reaching over 10\% in the most active accretors (Calvet \& Gullbring ~\cite{Calvet98}; Johns-Krull \& Gafford~\cite{J-Krull02}, Ingleby et al.~\cite{Ingleby13}).

S Cr A SE is a heavy accretor, and by assuming a filling factor of 5\% ($w$ = 0.05) and $\dot{M}_{acc}/w$ = 10$^{-6}$ M$_\odot$/yr, the mass accretion rate is 

   $$   \dot{M}_{acc} \approx 0.5\cdot~10^{-7} M_\odot/yr.  $$ 
	
For a more precise estimate the size of the hot spot needs to be found from modeling the accretion continuum from a broadband spectral energy distribution.  

The calculated accretion rate is consistent with that derived from the empirical relationship between the accretion rate and the H and He emission line luminosities (e.g., Fang et al.~\cite{Fang09}; Rigliaco et al.~\cite{Rigliaco12}). The H$\alpha$ line luminosity yields an  accretion rate of (0.3--0.9)~$\cdot$~10$^{-7}$~M$_\odot$/yr, while the \ion{He}{1} $\lambda5875$ line luminosity yields (1--2)~$\cdot$~10$^{-7}$~M$_\odot$/yr.   

\section{Discussion and conclusions}
\label{sec:conclusions}

The results presented in Fig.~\ref{ne_te} show that both the electron density and the temperature of the accreting gas increase closer to the star. The temperature distribution along the accretion channel is one of the critical parameter of the magnetospheric  models. Different heating mechanisms have been considered, including radiation from the hot material near the base of the accretion column, adiabatic compression of gas in the converging magnetic funnel, and generation and dissipation of MHD waves (Hartmann et al.~\cite{Hartmann94}; Martin~\cite{Martin96}; Muzerolle et al.~\cite{Muzerolle01}).

In the model of Muzerolle et al.~(\cite{Muzerolle01}), the temperature is relatively constant (7000--7500 K) throughout most of the magnetosphere. In the model of Martin~(\cite{Martin96}), the temperature increases toward the star, reaching 6000--7000 K near the stellar surface, which is more consistent with our results. However, the model stellar parameters in the Martin model ($M_*$ = 0.8 M$_\odot$, $R_*$ = 2 R$_\odot$) are somewhat different from those for S CrA SE. We note that the increase in temperature and degree of  ionization toward higher infall velocities is obvious already from the observed inverse P Cygni profiles, without any  special assumption.

The particle densities derived for S~CrA SE ($N_{\rm H}$ $\approx$ 10$^{13.5}$ cm$^{-3}$) are several times higher than in the models mentioned above. This may be related to the geometry of magnetic field. The traditional models assume a dipole configuration of magnetosphere. More complex structures of magnetic fields in CTTs were found from numerical simulations of accretion flows (see Romanova et al.~(\cite{Romanova11}, \cite{Romanova13}, and references therein) and from observations (e.g., Gregory \& Donati~\cite{Gregory11}). For a dipole plus octupole configuration the density in the accretion channel is higher and increases steeper toward the star than in the case of a dipole field (Gregory et al. \cite{Gregory07}; Adams \& Gregory~\cite{Adams12}). 

To summarize, we have monitored the two components of the binary T Tauri star S CrA during four nights at high spectral resolution. Both stars have the same surface temperature, radius, and mass. The SE component shows broad redshifted absorption components that flank many emission lines of different elements. During one night, these features were particularly prominent, and from line ratios at different observed redshifts the structures of the accretion flow to the stellar surface was derived in terms of temperature and electron density. This is the first time such an analysis has been made. 

Additional monitoring is warranted of other YY Ori stars with pronounced inverse P Cygni profiles in lines of neutral and ionized metals to clarify whether the physical structure of the accretion funnels are similar to or different from that of S CrA SE. 

\begin{acknowledgements}
This work was supported by an INTAS grant, the Swedish National Space Board, the Magnus Bergvall foundation and L\"angmanska kulturfonden. HCS acknowledges grant 621-2009-4153 of the Swedish Research Council. We thank the referee Scott Gregory for valuable comments.
\end{acknowledgements}

\end{document}